\documentclass[twocolumn,showpacs,preprintnumbers,amsmath,amssymb]{revtex4}

\usepackage{amsmath,amssymb,bbold,graphics}

\usepackage{graphicx}% Include figure files
\usepackage{dcolumn}% Align table columns on decimal point
\usepackage{bm}% bold math

\newcommand{\beq}{\begin{equation}}
\newcommand{\eeq}{\end{equation}}
\newcommand{\bea}{\begin{array}}
\newcommand{\eea}{\end{array}}

\begin{document}

\preprint{UCRHEP-T373}
\preprint{hep-ph/0404084}

\title{
Large neutrino mixing and normal mass hierarchy:\\ a discrete understanding
}

\author{Shao-Long Chen}
 \email{shaolong@citrus.ucr.edu}
\author{Michele Frigerio}
 \email{frigerio@physics.ucr.edu}
\author{Ernest Ma}
 \email{ma@physics.ucr.edu}
\affiliation{
Department of Physics, University of California, Riverside, 
CA 92521, USA
}

\date{\today}

%%%%%%%%%%%%%%%%%%%%%%%%%%%%%%%%%%%%%%%%%%%%%%%%%%%%%%%%%%%%%%%%%%%%%%%%%%%%%
\begin{abstract}
We discuss the possibility of flavor symmetries to explain the pattern
of charged lepton and neutrino masses and mixing angles.
We emphasize what are the obstacles for the generation of an almost 
maximal atmospheric mixing and what are the minimal ingredients to obtain it. 
A model based on the discrete symmetry $S_3$ is constructed,
which leads to the dominant $\mu\tau$-block
in the neutrino mass matrix, 
thus predicting normal hierarchy. 
This symmetry makes it possible to reproduce current data and predicts 
$0.01\lesssim\theta_{13}\lesssim 0.03$ and 
strongly suppressed neutrinoless $2\beta$-decay.
Moreover, it implies a relation between lepton and quark mixing angles:
$\theta_{23}^q \approx 2(\pi/4-\theta_{23})$. 
The Cabibbo mixing can also be reproduced
and $\theta_{13}^q\sim \theta_{12}^q\theta_{23}^q$.
$S_3$ is thus a candidate to describe all the basic  features of 
Standard Model fermion masses and mixing.
\end{abstract}

\pacs{14.60.Pq, 12.15.Ff, 11.30.Hv, 12.60.Fr}

\maketitle

%%%%%%%%%%%%%%%%%%%%%%%%%%%%%%%%%%%%%%%%%%%%%%%%%%%%%%%%%%%%%%%%%%%%%%%%%%%%
\section{Introduction}
%%%%%%%%%%%%%%%%%%%%%%%%%%%%%%%%%%%%%%%%%%%%%%%%%%%%%%%%%%%%%%%%%%%%%%%%%%%%

In the last few years our knowledge of the flavor structure of leptons has 
been strongly improved, thanks to neutrino oscillation experiments 
\cite{skatmo,sksolar,sno,snosalt,kamland,chooz}.
The hierarchy between the solar and atmospheric mass squared differences is
given by a factor $\Delta m^2_{sol}/\Delta m^2_{atm}\approx 0.035$. 
This translates into a very mild hierarchy
between two mass eigenstates, $m_2/m_3 \gtrsim 0.15$. 
The mass $m_1$ can be much smaller or almost equal to the other two, depending
on the degree of degeneracy of the spectrum.
The corresponding hierarchy parameters for charged leptons are 
$m_\mu/m_\tau\approx 1/20$ and $m_e/m_\mu\approx 1/200$.
The mixing between second and third generation is almost maximal 
($\sin^2 2\theta_{23}\gtrsim 0.9$), that between the first and the second is
large but non-maximal ($\sin^2 2\theta_{12}\approx 0.8$) and finally
$\sin^2 2\theta_{13} \lesssim 0.15$.
Also in the quark sector the hierarchy 
between first and second generation 
masses is stronger than between second and third, but
the mixing pattern is reversed: $\theta_{23}^q$ is much smaller than the 
Cabibbo angle $\theta_{12}^q$.

In the search for the underlying flavor symmetry dictating the relations
among fermion masses and mixing, one has to understand what features of
the data are directly connected with the symmetries of the mass matrices 
and what are second order effects. In this paper we take the point of 
view that the first step should be to explain the almost maximal atmospheric 
mixing. This angle is at present poorly constrained ($37^\circ \lesssim 
\theta_{23} \lesssim 53^\circ$ at 90\% C.L.) and next generation experiments
will not improve this bound significantly \cite{LBL}. 
However, there are already many indications
that such a large mixing requires a specific mechanism for its generation,
being $\theta_{23}$ exactly maximal or not:
\begin{itemize}
\item Present data allow for several structures of the Majorana neutrino 
mass matrix $M_\nu$, depending on the mass spectrum and CP-violating phases. 
In most cases, the large atmospheric mixing is imprinted in the dominant
structure of the matrix, the unique exception being 
$M_\nu\approx m_o\mathbb{1}$ \cite{M1,M2}.
\item In the case of normal mass hierarchy, which is the one closer to other
fermion mass spectra, the neutrino mass matrix is dominated by the $\mu\mu$,
$\mu\tau$ and $\tau\tau$ entries \cite{vissF1}, 
because the heaviest mass eigenstate is
almost equally shared between $\nu_\mu$ and $\nu_\tau$.
\item Since neutrinos and charged leptons belong to the same $SU(2)_L$
representation, one should naively expect a flavor alignment between them
(that is a cancellation between the left-handed mixing in the two mass 
matrices). 
The alignment is observed, in fact, between down and up quarks ($\theta_{23}^q
\approx 2^\circ$). Notice that this argument is roughly independent 
from the Majorana nature of the neutrino mass matrix.
\item The mixing can be enhanced through Renormalization Group running 
from high energy to electroweak scale 
(see \cite{cein2,lindner4} and references therein).
This works only for quasi-degenerate neutrino masses and in general 
the enhancement is efficient only for the solar mixing. 
The unique exception is, once again, the case $M_\nu \approx m_0\mathbb{1}$,
but the pattern of radiative corrections has to be chosen {\it ad hoc} 
\cite{M3} or the initial conditions at the high scale have to be fine-tuned
\cite{raduni,CEN2}.  
\end{itemize} 

In the context of the seesaw mechanism for the generation of small
neutrino masses, the possibility of a dynamical origin of the large mixing 
has been investigated. In the case of type I seesaw \cite{yana,gla,GRS,mose},
conditions have been found for a 'seesaw enhancement' of lepton mixing
\cite{enha,enha2}.
Specific correlations between the neutrino Dirac mass matrix and 
the associated right-handed Majorana 
mass matrix are required. Recently it has been pointed out \cite{BSV} that, 
in unified models with dominant type II seesaw 
\cite{LSW,mose2,SV,masa}, 
maximal atmospheric mixing can be related phenomenologically
with $b-\tau$ Yukawa unification.
In both cases, it seems to us that an underlying symmetry is
still required to enforce $\theta_{23}$ to be almost maximal.
In particular, we will show that the resulting value of the mixing at low 
energy depends crucially on which heavy fields give the dominant contribution
to $M_\nu$ and what are their flavor symmetries.

Many flavor models for leptons have been developed based on discrete or 
continuous symmetries, both Abelian and non-Abelian. 
The most popular class of models is based on $U(1)$ flavor symmetries 
and the Froggatt-Nielsen mechanism \cite{fronie}.
A review of $U(1)$ flavor symmetries in the lepton sector and references
can be found in \cite{alfe2}. 
Other models make use of the Abelian discrete symmetries $Z_n$ 
\cite{FMTY,glZ2,refle,bgw,tripa}.
However, minimal realizations of Abelian symmetries encounter some 
difficulties in reproducing data.
In section \ref{crit} we will comment on the present status of 
$U(1)$ and $Z_n$ flavor symmetries
in confronting neutrino oscillation data 
and in particular the large atmospheric mixing (see also \cite{low}).
To overcome at least some of the problems of Abelian symmetries 
and obtain greater predictiveness, a variety of non-Abelian 
symmetries have been used, both discrete \cite{D75,fras,A4,D4} and continuous 
\cite{bero,BHKR,kr3}.

In section \ref{model} we will show
that the basic features of lepton masses and mixings (and also those of
quarks) can be traced back to a minimal realization
of the smallest non-Abelian group, $S_3$, i.e. the permutations of
three objects. This group was first  used for flavor
physics in \cite{PS78} 
and it has been analyzed, for example, in few other papers 
\cite{derma,perma,MPP,hs3,KMR}.
The less minimal possibility of $S_{3L}\times S_{3R}$ symmetry was
first considered in \cite{HHW} and subsequently exploited in 
\cite{koi,frpl,CHM,FTY,tanLMA}.
Motivations for the use of $S_3$ flavor symmetries in supersymmetric models
can be found in \cite{hamu}: in particular the SUSY flavor problem can be 
relaxed.

In our approach, the choice of $S_3$ is minimal since we have 
to deal with three generations of fermions and we need at least one 
$2$-dimensional irreducible representation (irrep), 
in order to connect the two
generations which maximally mix. The group $S_3$ has, in fact, three irreps:
${\bf 1},{\bf 1'}$ and ${\bf 2}$.
It turns out that the existence of two inequivalent 1-dimensional 
representations is crucial to reproduce fermion masses and mixing.
We will analyze first the lepton $2-3$ sector (section \ref{muta}), then
we will extend our model to include the first generation (section
\ref{elec}), finally we will consider the quark sector (section 
\ref{quark}). In section \ref{disc} we summarize our results and
discuss merits and limits of the model.

%%%%%%%%%%%%%%%%%%%%%%%%%%%%%%%%%%%%%%%%%%%%%%%%%%%%%%%%%%%%%%%%%%%%%%%%%%%%
\section{A critical view of Abelian flavor symmetries \label{crit}}
%%%%%%%%%%%%%%%%%%%%%%%%%%%%%%%%%%%%%%%%%%%%%%%%%%%%%%%%%%%%%%%%%%%%%%%%%%%%

Let us discuss first continuous Abelian symmetries.
The most often considered $U(1)$ charge in the lepton sector is
$L_e-L_\mu-L_\tau$, introduced long ago in connection with pseudo-Dirac
neutrinos \cite{lelmlt}. In fact, this non-standard lepton charge induces 
naturally large mixing. However, the predicted phenomenology  is no longer
compatible with present data, at least in minimal realizations of the
flavor symmetry. Let us review  the basic reasons for the tension
between $L_e-L_\mu-L_\tau$ and experiments.

Let us call $q_X$ the $U(1)$-charge of the field $X$.
In leading order, only the couplings with field charges adding up to zero 
are allowed \cite{fronie}. 
We denote with $(\nu_\alpha~l_\alpha)^T$ the isodoublet of left-handed 
Weyl spinors with charge $0$ and $-1$, respectively, and
with $l^c_\alpha$ the isosinglet partners, 
which are left-handed and have charge $+1$.
Since $q_{e,\nu_e}=1$ and $q_{\mu,\nu_\mu,\tau,\nu_\tau}=-1$, 
the matrices of $U(1)$-charges relevant for the 
neutrino and charged lepton mass matrices $M_\nu$ and $M_l$ are
\beq\bea{c}
Q_{\nu\nu}=\left(\bea{ccc} 2 & 0 & 0 \\ 0 & -2 & -2 \\ 0 & -2 & -2 \eea\right)
~,\\
Q_{ll^c}=\left(\bea{ccc} 1+q_{e^c} & 1+q_{\mu^c} & 1+q_{\tau^c} \\
-1+q_{e^c} & -1+q_{\mu^c} & -1+q_{\tau^c} \\
-1+q_{e^c} & -1+q_{\mu^c} & -1+q_{\tau^c} \eea\right) ~,
\label{chau}\eea\eeq
where the charges of $e^c,\mu^c,\tau^c$ are not yet assigned.

The structure of the mass matrices depends on $q_\phi$, where $\phi$ is the
Standard Model Higgs isodoublet. 
Only two viable neutrino mass matrices can be obtained via the usual
five-dimensional operator $\nu\nu\phi\phi$, for $q_\phi=0$ and
$q_\phi=1$ respectively:
\beq
M_\nu^I=\left(\bea{ccc} 0 & a & b \\ a & 0 & 0 \\ b & 0 & 0 
\eea\right) ~,~~~~~~
M_\nu^N=\left(\bea{ccc} 0 & 0 & 0 \\ 0 & c & d \\ 0 & d & e 
\eea\right) ~,
\eeq
where $a$ and $b$ ($c,d$ and $e$) are of the same order. 
It is easy to check that $M_\nu^I$
corresponds to inverted mass hierarchy with eigenvalues 
$0,\pm\sqrt{a^2+b^2}$, with an order one $23$-mixing ($\tan\theta = a/b$)
and a maximal mixing between the two mass degenerate states. The matrix
$M_\nu^N$ corresponds to normal mass hierarchy with one zero eigenvalue and
order one mixing between the two massive states.

As far as charged leptons are concerned, $q_{\tau^c}$ should be chosen
to allow a non-zero $33$-entry in $M_l$, in order to generate the dominant
$\tau$ mass. Then, it is straightforward to show that the structure of 
$M_lM_l^\dag$ is given by
\beq
M_lM_l^\dag = \left(\bea{ccc} A & 0 & 0 \\ 0 & B & C \\ 0 & C^* & D 
\eea\right) ~.
\eeq
The parameters $B,C,D$ are of the same order, while $A$ can be suppressed
by the choice of $q_{e^c}$ and $q_{\mu^c}$. In any case, the contribution
of charged leptons to the mixing amounts only to an order one $23$-mixing.

Therefore, in both normal and inverted hierarchy cases, 
the $23$-mixing can be large but it is not naturally maximal. Moreover, if
all order one parameters are taken to be really close to $1$ (thus leading 
to $m^2_\mu \ll m^2_\tau$ in $M_lM_l^\dag$ and to $\Delta m^2_{sol}\ll
\Delta m^2_{atm}$ in $M_\nu^N$), 
the $2-3$ mixings are almost maximal in the two sectors, but
they cancel each other almost completely!
The $12$-mixing is maximal in the case of inverted hierarchy and zero in the 
normal hierarchy case, both in disagreement with experiment. 
Symmetry breaking corrections to these predictions are usually small and
do not reproduce data easily.

The problem to generate maximal atmospheric mixing is common to all
minimal models with $U(1)$ flavor symmetry.
The assignment of the same charge to
$\mu$ and $\tau$ isodoublets leads to a cancellation between
large mixing in neutrino and charged lepton mass matrices.
Models with inverted hierarchy can hardly explain the deviation of solar
mixing from maximal. Models embedded in Grand Unification theories usually
favor normal hierarchy, but the smallness of first generation masses tend to
prevent a large mixing in the $12$-sector.

Let us discuss now  the Abelian discrete symmetries $Z_n$.
In this type of models the fields transform under $Z_n$ via 
discrete rotations, given
by $1,\omega,\omega^2,\dots,\omega^{n-1}$, where $\omega\equiv e^{2\pi i/n}$
is the n-th root of unity. A coupling among  a given set of fields 
is allowed by the symmetry only if the product of the corresponding rotation
phases is equal to one.

In the case of $Z_2$, one can assign muon and tau leptons to the 
representation with phase $\omega\equiv-1$ and assume that the electron 
leptons are $Z_2$ invariant.
It is easy to check that the same problems are found as in the case of
$L_e-L_\mu-L_\tau$ symmetry: order one $2-3$ mixings are generated in both
neutrino and charged lepton sector and they tend to cancel each other; 
moreover the other two mixings are zero in the limit of exact $Z_2$ symmetry.

The situation is better in the case of $Z_3$. One can assume $e,\mu,\tau$ 
isodoublets to transform as $1,\omega,\omega^2$ respectively, where
$\omega\equiv e^{2\pi i/3}$. Then the analog of Eq.(\ref{chau}) is
\beq\bea{c}
\nu\nu\phi\phi \sim 
\left(\bea{ccc} 1 & \omega & \omega^2 \\ \omega & \omega^2 & 1 \\ 
\omega^2 & 1 & \omega \eea\right)a^2_\phi
~,\\
l l^c \phi \sim 
\left(\bea{ccc} a_{e^c} & a_{\mu^c} & a_{\tau^c} \\
\omega a_{e^c} & \omega a_{\mu^c} & \omega a_{\tau^c} \\
\omega^2 a_{e^c} & \omega^2 a_{\mu^c} & \omega^2 a_{\tau^c}
\eea\right)a_\phi ~,
\label{chaz}\eea\eeq
where $a_{e^c,\mu^c,\tau^c,\phi}$ are $1,\omega$ or $\omega^2$ 
depending on the $Z_3$ assignment of $e^c,\mu^c,\tau^c$ and $\phi$.
Whatever the assignment of these fields, 
it is clear that only
one element in each column of the charged lepton mass matrix is allowed.
This means that this matrix is diagonal up to an unobservable permutation
of the fields $e^c,\mu^c,\tau^c$. Therefore all the mixing comes from the 
neutrino sector and the only viable dominant structure is
\beq
M_\nu=\left(\bea{ccc} a & 0 & 0 \\ 0 & 0 & b \\ 0 & b & 0 \eea\right) ~,
\label{a4}\eeq
where $a$ and $b$ are of the same order. A maximal mixing
in the atmospheric sector is generated and a large solar mixing can appear
easily from subleading corrections. However, the neutrino spectrum has a very
unpleasant feature: in leading order the atmospheric mass difference 
is zero while the solar one is not. 

This drawback is generic to all models where the large atmospheric
mixing is obtained via dominant off-diagonal entries in the $2-3$ sector
of the neutrino mass matrix. In fact, phenomenology tells us that the two
maximally mixed states are associated with the largest mass splitting.
There are two ways to go round this difficulty. The first is to consider 
models predicting at leading order three degenerate neutrinos. In this case
both mass splittings are considered small perturbations while the maximal
$2-3$ mixing is an outcome of the symmetry. A good example is the $A_4$ model
\cite{A4}, which predicts the matrix (\ref{a4}) with $a=b$. The second way is
to construct the maximal mixing without dominant off-diagonal entries
in the $2-3$ sector of $M_\nu$. The advantage is that both mass differences 
are naturally
non-zero. We will pursue in the following this second way.

%%%%%%%%%%%%%%%%%%%%%%%%%%%%%%%%%%%%%%%%%%%%%%%%%%%%%%%%%%%%%%%%%%%%%%%%%%%%
\section{The $S_3$ model \label{model}}
%%%%%%%%%%%%%%%%%%%%%%%%%%%%%%%%%%%%%%%%%%%%%%%%%%%%%%%%%%%%%%%%%%%%%%%%%%%%

The $S_3$ group has six elements divided in three conjugacy classes:
the identity ($e$), the cyclic and anti-cyclic permutations of 
three objects ($g_c$ and $g_a$), the three interchanges of two objects 
leaving the third fixed ($g_1,~g_2,~g_3$).
Two independent 1-dimensional irreps are possible, depending if the action
of all six elements is trivial (${\bf 1}$) or if $g_i~(i=1,2,3)$ act
with a change of sign (${\bf 1'}$).
We will call ``odd'' $S_3$ singlets the fields transforming in the ${\bf 1'}$
representation.
The third and last irrep is 2-dimensional (${\bf 2}$). Since we deal only
with complex fields, we have the freedom to choose 
a complex realization of the
${\bf 2}$ \cite{DGP}, which leads to very convenient tensor product rules:
\beq\bea{cc}
R_2(e)=\left(\bea{cc} 1 & 0 \\ 0 & 1 \eea\right)~,~~~ &
R_2(g_c)=\left(\bea{cc} \omega & 0 \\ 0 & \omega^2 \eea\right)~,\\
R_2(g_a)=\left(\bea{cc} \omega^2 & 0 \\ 0 & \omega \eea\right)~,~~~ &
R_2(g_1)=\left(\bea{cc} 0 & \omega^2 \\ \omega & 0 \eea\right)~,\\
R_2(g_2)=\left(\bea{cc} 0 & \omega \\ \omega^2 & 0 \eea\right)~,~~~ &
R_2(g_3)=\left(\bea{cc} 0 & 1 \\ 1 & 0 \eea\right)~,
\eea\eeq
where $\omega\equiv e^{2i\pi/3}$. Notice that, in this realization,
\beq
\left(\bea{c}\psi_1\\\psi_2\eea\right)\in{\bf 2}
~~~\Rightarrow~~~
\left(\bea{c}\psi_2^\dag\\\psi_1^\dag \eea\right)\in{\bf 2}~.
\eeq
It is trivial to show that ${\bf 1}\times{\bf 1'}={\bf 1'}$, ${\bf 1'}\times
{\bf 1'}={\bf 1}$, ${\bf 2}\times{\bf 1}={\bf 2}$ and ${\bf 2}\times{\bf 1'}
={\bf 2}$. Finally and most importantly, one has
${\bf 2}\times{\bf 2}={\bf 1}+{\bf 1'}+
{\bf 2}$, where, if $(\psi_1~\psi_2)^T$ and $(\varphi_1~\varphi_2)^T$ are
$S_3$ doublets, then
\beq\bea{l}
\left(\bea{c}\psi_2\varphi_2\\\psi_1\varphi_1\eea\right)~,~~~
\left(\bea{c}\varphi_1^\dag\varphi_2
\\\varphi_2^\dag\varphi_1\eea\right) ~~~\in{\bf 2}~,\\
\psi_1\varphi_2+\psi_2\varphi_1~,~~~
\psi_1^\dag\varphi_1+\psi_2^\dag\varphi_2 ~~~\in{\bf 1}~,\\
\psi_1\varphi_2-\psi_2\varphi_1~,~~~
\psi_1^\dag\varphi_1-\psi_2^\dag\varphi_2 ~~~\in{\bf 1'}~.
\eea\eeq
With these few ingredients one can construct easily $S_3$ invariants,
once the assignment of Standard Model fields to $S_3$ irreps is given.

%%%%%%%%%%%%%%%%%%%%%%%%%%%%%%%%%%%%%%%%%%%%%%
\subsection{The $\mu\tau$ sector \label{muta}}
%%%%%%%%%%%%%%%%%%%%%%%%%%%%%%%%%%%%%%%%%%%%%%
 
Let the following fields transform under the $\bf{2}$ irrep of $S_3$:
\beq 
\left( \bea{c} L_\mu \\ L_\tau \eea \right) ~,~~~~~
\left( \bea{c} \Phi_1 \\ \Phi_2 \eea \right) ~,~~~~~
\left( \bea{c} \xi_1 \\ \xi_2 \eea \right) ~,
\eeq
where $L_\alpha=(\nu_\alpha~l_\alpha)^T$, $\Phi_i=(\phi^0_i~\phi^-_i)^T$
and $\xi_i=(\xi^{++}_i~\xi^+_i~\xi_i^0)^T$ are scalar isotriplets.
Let us assign also
\beq
\mu^c\in {\bf 1} ~,~~~~~ \tau^c\in {\bf 1'} ~.
\eeq
The invariants relevant for lepton masses are
\beq
(\tau\phi_1^0 + \mu\phi_2^0)\mu^c ~,~~~~~
(\tau\phi_1^0 - \mu\phi_2^0)\tau^c ~,~~~~~
\nu_\mu\nu_\mu\xi_1^0 + \nu_\tau\nu_\tau\xi_2^0 ~.
\label{inv1}\eeq
After electroweak symmetry breaking, the neutral scalar fields take
VEVs $<\phi_i^0>=v_i$ and $<\xi_i^0>=u_i$, so that
\beq\bea{rcl}
M_l &=& \left(\bea{cc} f_1v_2 & -f_2v_2 \\ f_1v_1  & f_2v_1 \eea\right) =\\
&=& \dfrac 12 \left(\bea{cc} 1 & -1 \\ 1 & 1 \eea\right)  
\left(\bea{cc} f_1(v_2+v_1) & f_2(v_1-v_2) \\ 
f_1(v_1-v_2) & f_2(v_2+v_1) \eea\right)  ~, \\
M_\nu &=& \left(\bea{cc} f_3u_1 & 0 \\ 0 & f_3u_2 \eea\right) ~,
\eea\eeq
where $f_i$ are dimensionless coupling constants.

It is apparent that, if $v_1=v_2=v$ (and $u_1\ne u_2$), 
maximal mixing is generated.
Indeed the condition $v_1=v_2$ minimizes the $S_3$ invariant scalar potential
(see section \ref{sp1}). The
lepton masses are given by $m_\mu=\sqrt{2}f_1 v$, $m_\tau=\sqrt{2}f_2 v$,
$\Delta m^2_{atm} = f_3^2(u_2^2-|u_1|^2)$ (notice that all complex phases
can be rotated away but a Majorana phase in the neutrino sector, which 
we can think as associated to $u_1$). The deviation from maximal
mixing can be easily computed as
\beq
\theta_{23}-\dfrac{\pi}{4} \approx \dfrac{v_1-v_2}{v_1+v_2} ~.
\eeq

A comment is in order about different contributions to the neutrino
mass matrix.
Since the VEVs of $\xi_i$ have to be seesaw suppressed 
(see section \ref{sp1}), in general a comparable
contribution to neutrino masses can come from the non-renormalizable operator
$L_\alpha L_\beta \bar{\Phi}_i \bar{\Phi}_j / M_R$, where $M_R$ is the seesaw
scale and $\bar{\Phi}_i\equiv i\sigma_2\Phi_i^* = 
(\phi_i^+ ~ -\phi_i^{0*})^T$. Taking into account that $(\bar{\Phi}_2 ~
\bar{\Phi}_1)^T\in \bf{2}$, one finds the following $S_3$ invariants:
\beq
\nu_\tau\nu_\tau(\phi_2^{0*})^2+\nu_\mu\nu_\mu(\phi_1^{0*})^2 ~,~~~~~~
(\nu_\mu\nu_\tau+\nu_\tau\nu_\mu)(\phi_1^0\phi_2^0+\phi_2^0\phi_1^0)^* ~.
\eeq
The first invariant can be mediated, at the seesaw scale, by the triplets 
$\xi_i$ and its contribution to neutrino masses modifies the values of $u_i$
but does not affect maximal mixing. The second invariant, on the contrary,
generates a non-zero off-diagonal entry in the neutrino sector which modifies
the resulting value of $\theta_{23}$, potentially preventing 
an almost maximal mixing. However, this invariant cannot be 
mediated by $\xi_{i}$, so in our minimal model it does not contribute.
One can check that it is mediated by scalar isotriplets $\xi\in \bf{1,1'}$ 
and/or by heavy neutrino states
$\nu^c\in\bf{2,1,1'}$. Maximal mixing indicates that all these fields are
absent or their contribution to $M_\nu$ is suppressed.

\subsubsection{Remarks on the scalar VEVs (I)\label{sp1}}
%%%%%%%%%%%%%%%%%%%%%%%%%%%%%%%%%%%%%%%%%%%%%%%%%%%%%%%%%

The most general $S_3$ invariant scalar potential for $\Phi_{1,2}$ is given by
\beq\bea{c}
V_\Phi = m^2(\Phi_1^\dag\Phi_1+\Phi_2^\dag\Phi_2) +
\dfrac{\lambda_1}{2}(\Phi_1^\dag\Phi_1+\Phi_2^\dag\Phi_2)^2 +\\
+\dfrac{\lambda_2}{2}(\Phi_1^\dag\Phi_1-\Phi_2^\dag\Phi_2)^2 +
\lambda_3 \Phi_1^\dag\Phi_2\Phi_2^\dag\Phi_1 ~.
\label{pot}\eea\eeq
Replacing $\Phi_{1,2}^{(\dag)}$ with $v_{1,2}^{(*)}$, one can check that
$V_\Phi(v_1,v_2)$ is bounded from below if and only if $\lambda_1+\lambda_2 >0$
and $-2\lambda_1<\lambda_3<2\lambda_2$. In this region of parameters,
the absolute minimum, in the case $m^2<0$, is given by $v_1=v_2=
-m^2/(2\lambda_1+\lambda_3)$. This justifies the assumption $v_1=v_2$ made
in the previous section.

Notice that $V_\Phi$ is invariant under two independent $U(1)$ 
transformations: $\Phi_{1,2}\rightarrow e^{i\theta_{1,2}}\Phi_{1,2}$.
As a consequence, electroweak symmetry breaking leaves us with an 
undesired real massless scalar, the residual Goldstone boson.
However, since the $S_3$ symmetry is broken at electroweak scale, one
can allow in $V_\Phi$ soft breaking terms with size comparable to the
quadratic term in Eq.(\ref{pot}). This $S_3$ explicit breaking can originate 
from the VEVs of extra fields belonging to a hidden sector of the theory,
similarly to the soft breaking terms in supersymmetric models.
Let us consider the extra term 
\beq
\Delta V_\Phi = \eta^2(\Phi^\dag_1\Phi_2 + \Phi_2^\dag \Phi_1) ~. 
\label{dpot}\eeq
It breaks a $U(1)$ symmetry since it enforces the relation 
$\theta_1=\theta_2$. In fact,
the term in Eq.(\ref{dpot}) respects the discrete symmetry $\Phi_1 
\leftrightarrow \Phi_2$, which is also a symmetry of the $S_3$ invariant
potential $V_\Phi$. Therefore the potential obtained adding 
Eq.(\ref{pot}) and Eq.(\ref{dpot})
can be still minimized by $v_1=v_2=v$, where now
$v=-(m^2+\eta^2)/(2\lambda_1+\lambda_3)$.
We assume that this ``custodial'' symmetry is preserved
till scales much smaller than electroweak.
Computing the quadratic part of $V_\Phi+\Delta V_\Phi$ after
electroweak symmetry breaking, one can check that all
physical scalar fields take a mass of the order of the electroweak scale.

The VEVs of $\xi_i$ are induced by the $\Phi_i$ VEVs via the following
scalar potential:
\beq
V_\xi = \dfrac 12 (M_\xi^2)^{ij}\xi_i^\dag\xi_j + (M_{\xi\phi\phi})^{ijk}
\xi_i\Phi_j\Phi_k + {\rm h.c.} ~.
\eeq
The $S_3$ invariant mass term, $M_\xi^2(\xi_1^\dag\xi_1 + \xi_2^\dag\xi_2)$,
is supposed to be very heavy, $M_\xi^2 \gg v^2$, in order to suppress the
triplet VEVs $u_i$ via the usual type II seesaw mechanism 
\cite{LSW,mose2,SV,masa}.
The actual values of $u_1$ and $u_2$ depend on the scale of different 
trilinear couplings. The $S_3$ invariant trilinear term is given by
\beq
M_{\xi\phi\phi}(\xi_1\Phi_1\Phi_1 + \xi_2\Phi_2\Phi_2)  + {\rm h.c.}~.
\label{dont}\eeq
If $M_{\xi\phi\phi}$ is the dominant trilinear coupling, then 
integrating out the heavy fields $\xi_i$ one obtains
$u_i^* = - v_i^2 M_{\xi\phi\phi}/M_\xi^2$, so that $v_1=v_2$ implies $u_1=u_2$.
However, if we assume that all the soft 
breaking term couplings are smaller than or equal to the electroweak scale, 
where $S_3$ is spontaneously 
broken, than there is no reason to expect that the
$S_3$ invariant trilinear coupling is dominating over the others.
For example, if the dominant trilinear term is
\beq
\sum_{i=1,2}M^i_{\xi\phi\phi}(\xi_i\Phi_1\Phi_2)  + {\rm h.c.}~,
\label{do}\eeq
than one obtains $u_i^*=
- v_1v_2 M^i_{\xi\phi\phi}/M_\xi^2$, so that $u_1\ne u_2$ is naturally induced.
Notice that Eq.(\ref{do}) preserves the discrete symmetry 
$\Phi_1 \leftrightarrow \Phi_2$. If this `custodial' symmetry is broken
only at scales much smaller than $v$, than it is natural to take
$M_{\xi\Phi\Phi}^i \sim v \gg M_{\xi\Phi\Phi}$.

Notice that the most general $V_\xi$ breaks $S_3$ only softly, 
thus not affecting
the mass matrices found in the previous section and allowing, 
at the same time, $u_1\ne u_2$.
The symmetry $\Phi_1\leftrightarrow\Phi_2$, that preserves $v_1=v_2$,
is somewhat analogue to the strong isospin, which is a good approximate
symmetry at the scale $\Lambda_{QCD}$ ($m_p \approx m_n$), not because 
$m_u\approx m_d$, but because $m_u,m_d \ll \Lambda_{QCD}$.

%%%%%%%%%%%%%%%%%%%%%%%%%%%%%%%%%%%%%%%%%%%%%
\subsection{The electron sector \label{elec}}
%%%%%%%%%%%%%%%%%%%%%%%%%%%%%%%%%%%%%%%%%%%%%

Let us introduce the fields
\beq
L_e~,~e^c~,~\Phi_3 \in \bf{1} ~.
\eeq
The $S_3$ singlet scalar isodoublet $\Phi_3$ is necessary to provide a 
non-zero mass to the electron.
The new invariants relevant for lepton masses are
\beq
(\tau\phi_1^0+\mu\phi_2^0)e^c ~,~~~~~
e e^c \phi_3^0 ~,~~~~~ e\mu^c \phi_3^0 ~,~~~~~
(\nu_\tau\xi_1^0+\nu_\mu\xi_2^0)\nu_e ~.
\label{inv2}\eeq
Comparing the first invariant in Eq.(\ref{inv1}) with the first in
Eq.(\ref{inv2}),
one realizes that only one linear combination of $\mu^c$ and $e^c$ 
is coupled to $(\tau\phi_1^0+\mu\phi_2^0)$, while the orthogonal is not.
Since a rotation of $\mu^c$ and $e^c$ is unobservable (right-handed),
we have the freedom to redefine $e^c$ as the decoupled state. Then 
the charged lepton mass matrix takes the form
\beq\bea{l}
M_l = \left(\bea{ccc} f_4 v_3 & f_5 v_3 & 0 \\
0 & f_1 v_2 & -f_2 v_2 \\
0 & f_1 v_1 & f_2 v_1 \eea\right) = \\
= \left(\bea{ccc} 1 & 0 & 0 \\
0 & \dfrac{1}{\sqrt{2}} & -\dfrac{1}{\sqrt{2}} \\
0 & \dfrac{1}{\sqrt{2}} & \dfrac{1}{\sqrt{2}} \eea\right)
\left(\bea{ccc} f_4 v_3 & f_5 v_3 & 0 \\
0 & \sqrt{2} f_1 v & 0 \\
0 & 0 & \sqrt{2} f_2 v \eea\right)
~,
\label{mll}\eea\eeq
where $v_3=<\phi_3^0>$ and in the last equality we have used 
$v_1=v_2=v$. 
Assuming $|v|\gg|v_3|$ (see section \ref{sp2}), one gets
\beq
\dfrac{m_e}{m_\mu} \approx \dfrac{|f_4v_3|}{\sqrt{2}|f_1 v|} ~,~~~~~~
\theta_{12}^l\approx \dfrac {|f_5 v_3|}{\sqrt{2}|f_1 v|} ~.
\label{12c}\eeq
If the coefficients $f_i$ are of order one, then $\theta_{12}^l
\sim m_e/m_\mu$.

The neutrino mass matrix has the following form:
\beq
M_\nu=\left(\bea{ccc} 0 & f_6 u_2 & f_6 u_1 \\
f_6 u_2 & f_3 u_1 & 0 \\
f_6 u_1 & 0 & f_3 u_2 \eea\right) =
f_3 u_2 \left(\bea{ccc} 0 & \epsilon_f & \epsilon_f \epsilon_u \\
\epsilon_f & \epsilon_u & 0 \\
\epsilon_f \epsilon_u  & 0 & 1 \eea\right) ~,
\label{massnu}\eeq
where $\epsilon_f\equiv f_6/f_3$ and $\epsilon_u\equiv u_1/u_2$.
Neglecting for the moment the small angle $\theta_{12}^l$,
the symmetry basis is given 
by $(\nu_e,~(\nu_\mu-\nu_\tau)/\sqrt{2},~(\nu_\mu+\nu_\tau)/\sqrt{2})$.
In flavor basis, one can write
\beq\bea{l}
M_\nu^{fl} = \dfrac{f_3 u_2}{2} \left[\left(\bea{ccc} 0&0&0 \\
0&1&1 \\ 0&1&1 \eea\right) + \right. \\
\left. + 
\left(\bea{ccc} 0 & \sqrt{2}\epsilon_f(1+\epsilon_u) & -\sqrt{2}\epsilon_f
(1-\epsilon_u) \\ \sqrt{2}\epsilon_f(1+\epsilon_u) & \epsilon_u & 
-\epsilon_u \\ -\sqrt{2}\epsilon_f(1-\epsilon_u) & -\epsilon_u & \epsilon_u
\eea\right)\right] ~.
\label{mnf}\eea\eeq
The $11$-entry in $M^{fl}_{\nu}$ is zero, thus
implying a strong suppression of neutrinoless 
$2\beta$-decay. It is well known that this can happen only in the
case of normal hierarchical neutrino spectrum. Therefore the parameters
$\epsilon_{f,u}$ have to be taken small and the first term in Eq.(\ref{mnf})
is the dominant $\mu\tau$-block \cite{vissF1}.
This dominant structure of $M_\nu^{fl}$ is the unique one allowed 
by data in the case of
normal mass hierarchy \cite{M1}. We have shown that
an $S_3$ symmetry is suitable to generate simply the $\mu\tau$-block.

Diagonalizing Eq.(\ref{massnu}), one finds
\beq\bea{l}
m_3\approx |f_3 u_2| \approx \sqrt{\Delta m^2_{atm}} ~,\\
\left|\theta_{23}-\dfrac{\pi}{4}\right|\approx|\epsilon_f^2\epsilon_u|
~,~~~~~~
\theta_{13} \approx |\epsilon_f\epsilon_u| ~,~~~~~~
\delta\approx \arg{\epsilon_u} ~,\\
\tan2\theta_{12}\approx 2\left|\dfrac{\epsilon_f}{\epsilon_u}\right| ~,~~~~~~
\dfrac{\Delta m^2_{sol}}{\Delta m^2_{atm}}\approx
\sqrt{|\epsilon_u|^4+4|\epsilon_u\epsilon_f|^2} ~,
\eea\label{pred}\eeq
where $\delta$ is the Dirac-type CP-violating phase in the standard
parameterization of the lepton mixing matrix \cite{PDG}.
The correlations among different observables are in agreement with present
data and can be tested in future precision measurements. Using the best fit
values $\tan2\theta_{12}=2.1$ and  
$\Delta m^2_{sol}/\Delta m^2_{atm}=0.035$ 
\cite{skatmo,sksolar,sno,snosalt,kamland}, we find 
$|\epsilon_u|\approx 0.12$ and $|\epsilon_f|\approx 0.13$, which imply
$|\theta_{23}-\pi/4|\approx 0.002$ and $\theta_{13}\approx 0.016$.

\begin{figure}
\includegraphics[width=246pt]{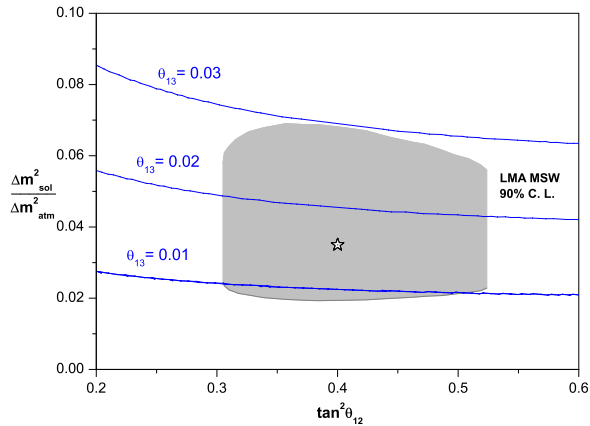}
\caption
{\label{fig}
Predictions of the $S_3$ model for the mixing angle $\theta_{13}$.
The gray area represents the Large Mixing Angle MSW allowed region for the
parameters $\tan^2\theta_{12}$ and $\Delta m^2_{sol}$, 
stretched to include also the
experimental uncertainty in $\Delta m^2_{atm}$. The best fit point is
denoted with a star. The values of $\theta_{13}$ are 
determined by $\theta_{12}$
and $\Delta m^2_{sol}/\Delta m^2_{atm}$ as described approximately by
Eq.(\ref{rela}). Possibly large corrections to these values of 
$\theta_{13}$ can come from charged lepton sector (see Eq.(\ref{male})).
}
\end{figure}

The allowed values of $\theta_{13}$ can be better evaluated noticing that
Eq.(\ref{pred}) implies
\beq
\theta_{13}\approx\dfrac 12 \sin 2\theta_{12}
\dfrac{\Delta m^2_{sol}}{\Delta m^2_{atm}} ~.
\label{rela}\eeq
Using $90\%$ C.L. allowed ranges, we obtain the prediction
\beq 
0.008\lesssim \theta_{13} \lesssim 0.032 ~.
\label{range}\eeq
Correspondingly, the parameters $|\epsilon_f|$ and $|\epsilon_u|$ 
are constrained in the range $0.1\div 0.2$.
A numerical diagonalization of the matrix (\ref{massnu}) has been performed
and the resulting predictions for $\theta_{13}$ are shown in Fig.\ref{fig},
in good agreement with the approximations given 
in Eqs. (\ref{rela}) and (\ref{range}).

We have neglected till now the contribution of the $12$-mixing in the
charged lepton sector (see Eq.(\ref{12c})). Performing a careful
commutation of rotation matrices, we find that $\theta_{12}^l$ affects
all three observable mixing angles as follows:
\beq
\Delta^l \theta_{23}\approx \dfrac{(\theta_{12}^l)^2}{4} ~,~~~~~~
\Delta^l \theta_{13}\approx \dfrac{\theta_{12}^l}{\sqrt{2}} ~,~~~~~~
\Delta^l \theta_{12}\approx \dfrac{\theta_{12}^l}{\sqrt{2}} ~.
\label{male}\eeq
In the case of $13$-mixing, this correction can be important even for
$\theta_{12}^l$ as small as $m_e/m_\mu\approx 0.005$.
The rotation $\theta_{12}^l$ generates also
\beq
m_{ee}\equiv|(M_\nu^{fl})_{11}| \approx \sqrt{2\Delta m^2_{atm}}|\epsilon_f|
\theta_{12}^l \lesssim 10^{-2}{\rm eV}\cdot \theta_{12}^l ~,
\eeq
which induces a non-zero (but still quite suppressed) neutrinoless
$2\beta$-decay.

It is worthwhile to give a look to the contribution of $\Phi_3$ to the
five-dimensional operator $L_\alpha L_\beta \bar{\Phi}_i \bar{\Phi}_j / M_R$,
which can perturb the neutrino mass matrix (\ref{massnu}).
The possible $S_3$ invariants are
\beq\bea{l}
\nu_e\nu_e(\phi_3^{0*})^2 ~,~~~~~~
(\nu_\mu\nu_\tau+\nu_\tau\nu_\mu)(\phi_3^{0*})^2 ~,~~~~~\\
\nu_e(\nu_\mu\phi_1^{0*}-\nu_\tau\phi_2^{0*})\phi_3^{0*} ~,~~~~~~
(\nu_\mu\nu_\mu\phi_2^{0*}+\nu_\tau\nu_\tau\phi_1^{0*})\phi_3^{0*} ~.
\eea\label{inv4}\eeq
The first two invariants are not mediated by $\xi_{1,2}$ and therefore are
absent in the minimal model. The last two invariants can be mediated, but
their contribution can be absorbed into a redefinition of $u_{1,2}$ and
$f_{3,6}$ in Eq.(\ref{massnu}).

\subsubsection{Remarks on the scalar VEVs (II)\label{sp2}}
%%%%%%%%%%%%%%%%%%%%%%%%%%%%%%%%%%%%%%%%%%%%%%%%%%%%%%%%%%

Of course the introduction of the $S_3$ singlet $\Phi_3$ in our model
modifies the scalar potential discussed in section \ref{sp1}. The most general
$S_3$ invariant potential is
\beq\bea{c}
V_{3\Phi}  =  V_\Phi+m_3^2\Phi^\dag_3\Phi_3 + \dfrac{\lambda_4}{2}
(\Phi^\dag_3\Phi_3)^2 +  \\  + \lambda_5 (\Phi^\dag_3\Phi_3)
(\Phi^\dag_1\Phi_1+\Phi^\dag_2\Phi_2) +\lambda_6 \Phi_3^\dag 
(\Phi_1\Phi^\dag_1+\Phi_2\Phi^\dag_2) \Phi_3 + \\
+ \left[ \lambda_7 \Phi^\dag_3 \Phi_1 \Phi^\dag_3 \Phi_2 +
\lambda_8 \Phi^\dag_3 (\Phi_1\Phi^\dag_2\Phi_1+\Phi_2\Phi^\dag_1\Phi_2)
+ {\rm h.c.}\right] ~
\eea\label{potpot}\eeq 
where $V_{\phi}$ is given in Eq.(\ref{pot}).

We do not discuss, here, the general minimization problem (see, for instance,
\cite{S3pot}). For our purposes, it is enough to verify that a 
solution exists with
$|v_1|=|v_2|\equiv|v| \gg |v_3|$. Assuming that all the couplings 
$\lambda_i$ are of the same order and barring special cancellations 
among them, we find a minimum for
\beq
|v|^2 \approx -\dfrac{m^2}{2\lambda_1+\lambda_3} ~,~~~~~~
v_3 \approx -\dfrac{2\tilde{\lambda}_8 |v|^3}
{m_3^2+2|v|^2(\lambda_5+\lambda_6+\tilde{\lambda_7})} ~,
\eeq
where $\tilde{\lambda}_{7,8}$ are appropriate rephasing of $\lambda_{7,8}$,
determined by the $v_i$ complex phases. To satisfy the initial assumption
$|v_3|\ll |v|$ is enough to choose $m_3^2 \gg |v|^2$, so that
$|v_3| \approx 2|\tilde{\lambda}_8 v|(|v|^2/m_3^2)$.

The large parameter $m_3^2$
determines, in first approximation, the mass of the four real scalars
contained in $\Phi_3$.
Since the effect of $v_3$ can be safely treated as a small perturbation, 
all further considerations made in section \ref{sp1} about 
$\Phi_{1,2}$ and $\xi_{1,2}$ masses and VEVs are still valid.

%%%%%%%%%%%%%%%%%%%%%%%%%%%%%%%%%%%%%%%%%%%
\subsection{The quark sector \label{quark}}
%%%%%%%%%%%%%%%%%%%%%%%%%%%%%%%%%%%%%%%%%%%

Let us assign quark fields to $S_3$ representations in exact 
analogy with leptons:
\beq
\left(\bea{c} Q_2 \\ Q_3 \eea\right) \in {\bf 2} ~,~~~~~~
Q_1,~u^c,~c^c,~d^c,~s^c \in {\bf 1} ~,~~~~~~
b^c,~ t^c \in {\bf 1'} ~,
\eeq
where $Q_i = (u_i~d_i)^T$. 
As for leptons, the third generation isosinglets are odd $S_3$ singlets
while second generation ones are $S_3$ invariant. 
This turns out to be the origin of a sizable $1-2$ mixing and a suppressed
$1-3$ mixing.
It is straightforward to construct independent invariants
contributing to quark mass matrices. The result is
\beq\bea{c}
M_u = \left(\bea{ccc} g^u_3 v^*_3 & g^u_4 v^*_3 & 0  \\
0 & g^u_1 v_1^* & -g^u_2 v_1^* \\
0 & g^u_1 v_2^* & g^u_2 v_2^* \eea\right) ~,\\
M_d = \left(\bea{ccc} g^d_3 v_3 & g^d_4 v_3 & 0 \\
0 & g^d_1 v_2 & -g^d_2 v_2 \\
0 & g^d_1 v_1 & g^d_2 v_1 \eea\right) ~.
\label{mumd}\eea\eeq
The $S_3$ assignment of the quarks parallels that of the leptons,
as shown by the analogous structure of their mass matrices (\ref{mll})
and (\ref{mumd}).  
This makes our model
suitable for a possible embedding in a Grand Unification Theory.
It is interesting that, in a class of $SO(10)$ inspired models known
as ``lopsided'' \cite{abb,BBlo}, the almost maximal leptonic $2-3$ mixing 
originates in the charged lepton mass matrix as in the present case.
However, the pattern is different in the quark sector:
in our model the left-handed $2-3$ mixing cancels between down and up 
quark sectors; in ``lopsided'' models the large charged lepton mixing 
appears also in the down quark mass matrix, but on the 
right-handed side, therefore it does not show up in the $CKM$ matrix.

In the limit $|v_1|=|v_2|=|v|$, the maximal $2-3$ mixing 
cancels exactly between up and down matrices  and we get
$m_c/m_t\approx |g_1^u/g_2^u|$, $m_s/m_b\approx |g_1^d/g_2^d|$. 
The experimental value,
$\theta_{23}^q \approx 0.04$, can be explained by
small corrections to $|v_1|=|v_2|$, due to soft 
breaking and/or $\Phi_3$ contributions to the scalar potential 
in Eq.(\ref{potpot}).
In fact, one finds the interesting sum rule
\beq
\theta_{23}^q \approx 2\left(\dfrac{\pi}{4}-\theta_{23}\right)
\approx 2\dfrac{|v_2|-|v_1|}{|v_2|+|v_1|} ~.
\label{sumrule}\eeq
Larger deviations from maximal atmospheric mixing can be explained only
by a non-zero contribution to the mixing coming from the neutrino sector
(see discussion at the end of section \ref{muta}).

In analogy to the charged lepton sector, after the maximal $2-3$ rotation
the structure of the $1-2$ blocks in $M_d$ and in $M_u$ implies
\beq\bea{c}
\dfrac{m_d}{m_s}\approx \dfrac{|g^d_3 v_3|}{\sqrt{2}|g^d_1 v|}\sim 
\dfrac{1}{20} ~,~~~~~
\dfrac{m_u}{m_c}\approx \dfrac{|g^u_3 v_3|}{\sqrt{2}|g^u_1 v|}\sim 
\dfrac{1}{400} ~,\\
\theta_{12}^{d,u}\approx \dfrac{|g^{d,u}_4 v_3|}{\sqrt{2}|g^{d,u}_1 v|} ~. 
\label{12b}\eea\eeq
The CKM mixing matrix is given by
\beq\bea{rcl}
U_{CKM} & \equiv & U_{23}U_{13}U_{12} = \\
& = & (U_{23}^u U_{12}^u)^\dag(U_{23}^d U_{12}^d)
\equiv U_{12}^{u\dag} U_{23}^q U_{12}^d ~,
\eea\eeq
where $U_{23}^{u,d}$ are almost maximal $23$-rotations that cancel up to
the small angle given in Eq.(\ref{sumrule}), while $U_{12}^{u,d}$ are the
$12$-rotations quantified in Eq.(\ref{12b}).
From the commutation of $U_{12}^{u\dag}$ and $U_{23}^q$, a $13$-mixing
is generated: $\theta_{13}^q\approx \theta_{12}^u\theta_{23}^q$.
One has to fit $\theta_{13}^q\approx 0.004$ and the Cabibbo angle
$\theta_{12}^q\approx 0.22$, which results from the combination of
$U_{12}^{u\dag}$ and $U_{12}^d$. Looking at Eq.(\ref{12b}), one 
realizes that the fit is successful for $|v_3/v|\sim 0.1$ and the coefficients
$g_i^{u,d}$ of order one. However, a significant
suppression of $g_3^u$ is required to match the smallness of the up quark
mass. We will suggest an explanation for this suppression in the next section.

%%%%%%%%%%%%%%%%%%%%%%%%%%%%%%%%%%%%%%%%%%%%%%%%%%%%%%%%%%%%%%%%%%%%%%%%%%%%
\section{Discussion and conclusions \label{disc}}
%%%%%%%%%%%%%%%%%%%%%%%%%%%%%%%%%%%%%%%%%%%%%%%%%%%%%%%%%%%%%%%%%%%%%%%%%%%%

We have analyzed the problem of constructing a maximal $2-3$ mixing in the
lepton sector. As shown in section \ref{crit}, minimal models
with Abelian flavor symmetries give a defective description
of neutrino oscillation data. 
We have then constructed a model based on $S_3$ flavor symmetry
which generates naturally maximal $2-3$ mixing. 
It contains only Standard Model particles plus an enlarged scalar sector,
formed by three isodoublets at electroweak scale and two much 
heavier isotriplets.

Let us summarize the ingredients of the model:
\begin{itemize}
\item 
Second and third generation fermion isodoublets transform as an $S_3$ doublet.
\item Second generation fermion isosinglets are $S_3$ invariants
while the third ones transform as odd $S_3$ singlets.
\item The two scalar isodoublets which generate the $2-3$ block of
charged fermion mass matrices have the same VEV, while the one
giving mass to first generation fermions takes a much smaller VEV.
\item The neutrino mass matrix is generated by an $S_3$ doublet of heavy
scalar isotriplets, which have tiny and different VEVs.
\end{itemize}

As a consequence, maximal $2-3$ mixing is induced in the charged fermion
mass matrices, while the $2-3$ block is diagonal in the Majorana mass matrix
of neutrinos. Therefore, a maximal mixing results in the lepton sector,
whereas complete cancellation takes place between up and down quark mixing.
Small corrections in both sectors are allowed and they are correlated as in 
Eq.(\ref{sumrule}).
Extra deviation from maximal atmospheric mixing can appear
if heavy fields other than the two isotriplets give a subdominant 
contribution to the neutrino mass matrix.
The ratio between second and third generation masses is not determined by
the $S_3$ symmetry. However $S_3$ distinguishes the two corresponding
sets of couplings, which are of the type ${\bf 2}\times{\bf 2}\times{\bf 1}$
and ${\bf 2}\times{\bf 2}\times{\bf 1'}$ for second and third generation 
respectively. This suggests that the hierarchy between the two types 
can be induced by extra flavor structure to be added to our minimal model.

The spectrum of neutrinos is with normal hierarchy. 
In flavor basis the neutrino mass matrix has a dominant $\mu\tau$-block.
The $1-2$ mixing can be naturally of order one, thus
explaining the Large Mixing Angle MSW solution of the solar neutrino problem.
The $1-3$ mixing is correlated with solar parameters by Eq.(\ref{rela}) 
and turns out to be about $0.02$. 

The neutrinoless $2\beta$-decay is strongly
suppressed. 
If the recent claim \cite{kla4} of neutrinoless $2\beta$-decay were
confirmed, then $|(M_\nu)_{11}|\gtrsim 0.1$ eV and our model would 
be ruled out, unless the dominant mechanism
of the decay is not the exchange of the light Majorana neutrinos 
\cite{wayout}.
Notice that in our model the suppression of $(M_\nu)_{11}$ is induced by
the requirement to obtain an atmospheric mixing close to maximal.
In fact, one can check that, adding $\xi\in{\bf 1},{\bf 1'}$ and/or
$\nu^c\in{\bf 2},{\bf 1},{\bf 1'}$ to the model, a non-zero contribution to
$(M_\nu)_{11}$ is accompanied by a contribution of the same order to
$(M_\nu)_{23}$, which tends to cancel the maximal $2-3$ mixing coming from
charged leptons.

In the quark sector, a small $1-2$ mixing is 
generated naturally because $u$ and $c^c$ ($d$ and $s^c$)
are both $S_3$ invariants, thus allowing a sizable $12$-entry in the mass
matrices. One can reproduce easily the Cabibbo angle.
The different $S_3$ assignment of $t^c$ and $b^c$ (odd $S_3$ singlets)
suppresses the $1-3$ mixing in $M_d$ and $M_u$; the resulting CKM matrix
contains $\theta_{13}^q\sim\theta_{12}^q\theta_{23}^q$, 
in agreement with data.

First generation masses are suppressed by 
the small ratio of scalar VEVs $|v_3/v|$. This ratio cannot be too small 
since the Cabibbo angle is correspondingly suppressed. 
In particular the smallness of the $u$ quark (electron) mass indicates
an extra source of suppression. This can be easily obtained, for example,
introducing a $Z_2$ parity leaving all fields invariant but $u^c$
($e^c$), which is $Z_2$ odd. It is easy to check that, in the limit of
exact $Z_2$ symmetry, $m_u$ ($m_e$) is forbidden.

The model can be tested in the near future by
\begin{itemize}
\item precision measurements of neutrino oscillation parameters;
\item upper bounds on neutrino masses from cosmology, tritium $\beta$-decay
and neutrinoless $2\beta$-decay;
\item direct investigation of the scalar isodoublet sector at LHC;
\item flavor violating decays mediated by the scalars.
\end{itemize}
A detailed study of the phenomenological implications 
of the model is left for future work.

In conclusion, $S_3$ is the smallest flavor symmetry group which can explain 
in a minimal way the maximal atmospheric mixing. The required structure
in the lepton $2-3$ sector enforces in a straightforward way the whole
structure of three generation lepton and quark mass matrices. These
matrices are suitable to explain all current data.

\section*{Acknowledgments}
%%%%%%%%%%%%%%%%%%%%%%%%%%

This work has been partially supported by the U.S. Department of Energy
under Grant No. DE-FG03-94ER40837.

\bibliography{PRDpaper}

\end{document}